\documentclass{emulateapj}
\def\chan{\textit{Chandra}}

\def\xmm{\textit{XMM--Newton}}
\def\snr{G292.2--0.5}
\def\psr{J1119--6127}

\def\j1846{J1846--0258}
\def\b1509{B1509--58}

\begin{document}

\title{Unusual Pulsed X-ray Emission from the Young, High Magnetic \\ Field Pulsar 
PSR J1119--6127}
\author{M. E. Gonzalez\altaffilmark{1}, V. M. Kaspi\altaffilmark{1}, F. 
Camilo\altaffilmark{2}, B. M. Gaensler\altaffilmark{3}, and M. J. 
Pivovaroff\altaffilmark{4}} 
\altaffiltext{1}{Department of Physics, Rutherford Physics Building, 
McGill University, Montreal, QC H3A 2T8, Canada.} 
%\altaffiltext{2}{gonzalez@physics.mcgill.ca; NSERC PGS B fellow} 
%\altaffiltext{3}{Canada Research Chair, Steacie Fellow.}
\altaffiltext{2}{Columbia Astrophysics Laboratory, Columbia University, 550 West 
120th Street, New York, NY 10027.}
\altaffiltext{3}{Harvard-Smithsonian Centre for Astrophysics, 60 Garden Street, 
Cambridge, MA 02138.}
\altaffiltext{4}{Lawrence Livermore National Laboratory, P.O. Box 808, L-258, 
Livermore, CA 94550.}

\begin{abstract}
We present \xmm\ observations of the radio pulsar PSR \psr, which has
an inferred age of 1,700~yr and surface dipole magnetic field strength of 
4.1$\times$10$^{13}$~G. We report the first detection of pulsed X-ray 
emission from PSR \psr. In the 0.5--2.0 keV range,
the pulse profile shows a narrow peak with a very high pulsed fraction of
(74$\pm$14)\%. In the 2.0--10.0 keV range, the upper limit for the pulsed 
fraction is 28\% (99\% confidence). The pulsed emission
is well described by a thermal blackbody model with a temperature of 
$T^{\infty}$~= 2.4$^{+0.3}_{-0.2}$$\times$10$^{6}$~K and emitting radius 
of 3.4$^{+1.8}_{-0.3}$~km 
(at a distance of 8.4~kpc). Atmospheric models result in problematic
estimates for the distance/emitting area. PSR \psr\ is now the radio pulsar 
with smallest characteristic age from which thermal X-ray emission has been 
detected. The combined temporal and spectral characteristics of this
emission are unlike those of other radio pulsars detected at X-ray energies
and challenge current models of thermal emission from neutron stars.

\end{abstract}


\section{Introduction}
The most commonly studied neutron stars (NSs) are radio pulsars, 
generally thought to be powered by the loss of rotational kinetic 
energy due to magnetic braking. 
The thermal X-rays observed from some pulsars are thought
to be emitted at the surface and to originate from initial cooling 
or from polar caps reheated by magnetospheric 
processes (see, e.g., Kaspi, Roberts \& Harding 2004, for a review).
%\citep[see, e.g.,][for a review]{krh04}. 
At low X-ray energies, this thermal emission shows broad pulses of 
low amplitude. In contrast, highly modulated non-thermal X-rays 
showing narrow pulses are thought to arise from synchrotron emission 
in the magnetosphere \citep[e.g.,][]{cz99,ba02}. Thus, X-ray studies 
of radio pulsars are an excellent diagnostic of the physics governing 
NS emission.

PSR \psr\ was discovered in the Parkes Multibeam Pulsar Survey \citep{ckl+00}.
It has a spin period $P$ = 0.408~s $\equiv$ 1/$\nu$ and period derivative 
$\dot{P}$ = 4.1$\times$10$^{-12}$. The measured braking index for the pulsar
of $n$ = 2.91$\pm$0.05 ($\dot{\nu}$ $\propto$ $-$$\nu^{n}$) implies an upper
limit for the age of 1,700~yr, making it one of the youngest
pulsars known. The pulsar
has a spin-down luminosity of $\dot{E}$ = 4$\pi^2$$I$$\dot{P}$/$P^{3}$ = 
2.3$\times$10$^{36}$~erg~s$^{-1}$ (for a moment of inertia $I$ $\equiv$ 
10$^{45}$~g~cm$^{2}$) and an inferred surface dipole magnetic field 
strength of $B$ $\equiv$ 3.2$\times$10$^{19}$($P$$\dot{P}$)$^{1/2}$~G = 
4.1$\times$10$^{13}$~G. This value of $B$ is among the 
highest known in the radio pulsar population. PSR \psr\ powers a small 
($3''$$\times$$6''$) X-ray pulsar wind nebula \citep[PWN, ][]{gs03} and 
lies close to the center of the $15'$-diameter supernova remnant 
(SNR) \snr\ \citep{cgk+01b,pkc+01}. HI absorption measurements imply a 
kinematic distance for the remnant of 8.4$\pm$0.4~kpc \citep{cmc04}, in
agreement with estimates made from its location in the Carina spiral arm 
\citep{ckl+00}.

Here we report on \xmm\ observations of PSR \psr\ and SNR \snr. 
The X-ray emission from PSR \psr\ shows a high-temperature thermal 
spectrum  having a high pulsed fraction at low 
energies. No pulsations are detected at high energies. These characteristics 
are unlike those of other radio pulsars detected at X-ray energies. We 
discuss the possible origins of this emission. A detailed analysis of SNR 
\snr\ is deferred to a future paper.

\section{Observation and Imaging Analysis}
SNR \snr\ and PSR \psr\ were observed with \xmm\ on 2003 June 26. 
The European Photon Imaging Camera (EPIC)  instruments MOS 
and PN were operated in full-window and large-window mode,
respectively, using the medium filter for both. The temporal 
resolution was 2.6~s for MOS and 48~ms for PN. The data 
were analysed using the Science Analysis System software 
(SAS v6.2.0) and standard reduction techniques. The last 5~ks of the PN 
observation were rejected due to high radiation.
The resulting exposure time was 48~ks for MOS1/MOS2 and 43~ks for PN. 

\begin{figure}
\begin{center}
%\epsscale{0.75}
\centerline {\includegraphics{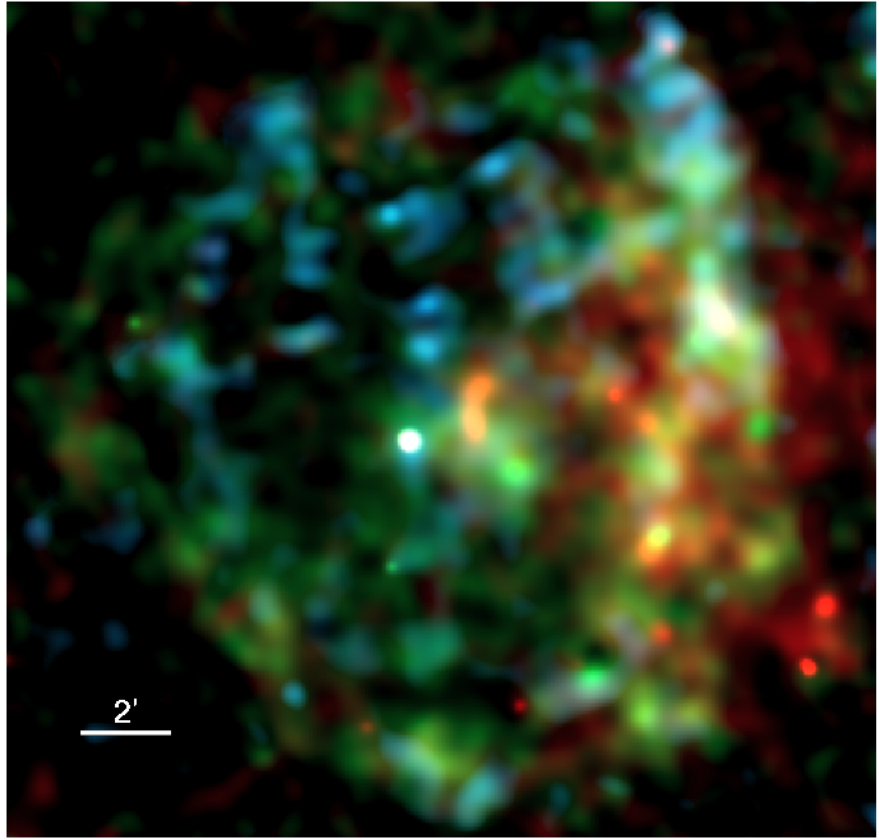}}
\caption{\label{FigMOS} EPIC-MOS combined image of SNR \snr\ and PSR \psr\ 
(19$'$$\times$19$'$ field) in the 0.3--1.5 keV ({\it red}), 1.5--3.0 keV 
({\it green}) and 3.0--10.0 keV ({\it blue}) ranges.
The image reveals for the first time the detailed
morphology of the remnant at X-ray energies.
The large east-west asymmetry at low energies has been attributed
to the presence of a molecular cloud on the east side of the 
field \citep{pkc+01}.}
\end{center}
\epsscale{1}
\end{figure}

%\section{Imaging Analysis}\label{SecImg}
Figure \ref{FigMOS} shows the combined MOS image of the 
system in the 0.3--1.5 keV ({\it red}), 1.5--3.0 keV ({\it green}) 
and 3.0--10.0 keV ({\it blue}) bands.
Individual MOS1/MOS2 images were first binned into pixels of 
2$\farcs$5$\times$2$\farcs$5 and other point sources in the field were 
excluded. These images were then added and adaptively 
smoothed with a Gaussian with $\sigma$=5$''$--15$''$ to obtain 
signal-to-noise ratios higher than 3$\sigma$. Background images and exposure 
maps at each energy band were similarly obtained and used to correct 
the final images.  The detailed spatial distribution in the 3.0--10.0 keV 
({\it blue}) image should be examined with caution as this 
energy band suffered from a high degree of stray-light contamination
on \xmm's mirrors from a nearby high-energy source. 
%used to background-subtract and flat-field the image, respectively.
Here we focus on the bright source at the center of Fig. \ref{FigMOS}, 
which has coordinates $\alpha_{2000}$=11$^{h}$19$^{m}$14$\fs$65 
and $\delta_{2000}$=$-$61$\degr$27$'$50$\farcs$2 (4$''$ error). This
position coincides with the \chan\ and radio coordinates of PSR \psr. 
The spatial resolution of \xmm\ (half-power diameter of 15$''$) 
does not allow for the arc-second scale PWN discovered with \chan\ to be
resolved.

\section{Timing Analysis}\label{SecTim}
The PN data were used to search for pulsations from PSR \psr. 
Examining the raw event list as in \cite{wkt+04} revealed an 
instrumental timing anomaly that resulted in time tags 
being shifted forward by 1 s for events detected after the anomaly took 
place. We corrected for this
by subtracting 1~s from the original time stamps and from 
the auxiliary PN file (*AUX.FIT) for the observation. The 
corrected dataset was then converted
to the solar system barycenter.

A circular region of 25$''$ radius was used to extract the pulsar events. 
The data were divided into 
different energy ranges at 0.5--10.0 keV (620$\pm$30 photons), 
0.5--2.0 keV (340$\pm$25 photons) and 2.0--10.0 keV (275$\pm$22 photons).
From an updated radio ephemeris obtained from regular monitoring at the
Parkes radio telescope, 
the predicted barycentric radio period for the middle of our observation 
(MJD 52816.54) was 2.44972914 Hz. 
The $\textit{Z}^{2}_{n}$ test \citep{bbb+83} with harmonics $n$~= 
\{1,2,4,8\} was used to search for pulsations.
The most significant signal was detected in the 0.5--2.0 keV range with 
$\textit{Z}^{2}_{2}$~= 52.8 (6.6$\sigma$ significance) at a frequency of 
2.449726(6) Hz (1$\sigma$ errors). This frequency is in agreement with 
the radio prediction for PSR \psr. 
In the 2.0--10.0 keV range, no signal was found with a significance 
$>$2.8$\sigma$. In the 0.5--10.0 keV 
range, the above signal was detected with a 5.2$\sigma$ significance.

\begin{figure}
\begin{center}
%\epsscale{0.75}
\centerline {\includegraphics{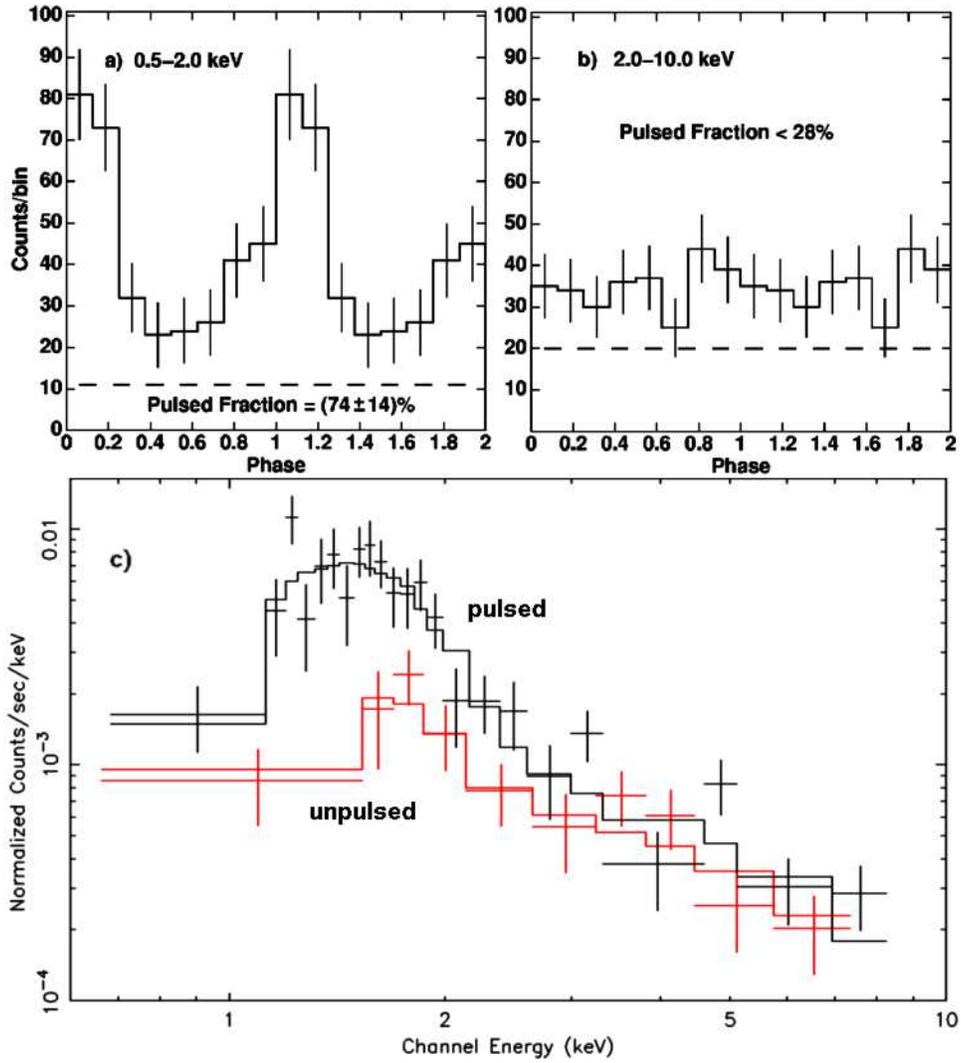}}
\caption{\label{FigProfPha} {\it Top}: X-ray pulse profiles of PSR \psr\
in the 0.5--2.0~keV ({\it left}) and 2.0--10.0 keV ({\it right}) ranges.
Errors bars are 1$\sigma$ and two cycles are shown. The peak of the radio
pulse is at phase 0. The dashed lines represent our estimates for the 
contribution from the pulsar's surroundings (see \S\ref{SecEmis}). 
{\it Bottom}: EPIC-PN spectra obtained for the pulsed 
($black$) and unpulsed ($red$) regions of the pulse profile with their 
respective best-fit blackbody plus power-law model (solid curves). {\it
[See the electronic version of the Journal for a color version of this
figure.]}}
\end{center}
\end{figure}

Figure \ref{FigProfPha} shows the resulting pulse 
profiles at 0.5--2.0 keV (top, left) and 2.0--10.0 keV (top, right). 
The background was estimated from a nearby region away from 
bright SNR knots. The horizontal dashed lines represent our estimates for 
the contribution from the pulsar's surroundings (see \S\ref{SecEmis}).
The resulting pulsed fraction [PF $\equiv$ 
($F_{max}$--$F_{min}$)/($F_{max}$+$F_{min}$)] is labeled in Fig. 
\ref{FigProfPha} (1$\sigma$ statistical errors). In the 2.0--10.0 keV 
range, we derive an upper limit for the pulsed fraction of 28\% (at the 99\%
confidence level). For this, we used the maximum power obtained from a
Fourier analysis of the data for a small range of frequencies centered 
on the radio prediction (within our errors) and 
assumed a worse-case sinusoidal profile. An upper limit for the pulse 
amplitude can then be derived by calculating an upper limit on the 
power that could still be present above the observed maximum at a 
specific level of confidence. For a detailed description of this 
method we refer to \cite{vvw+94}, and e.g., \cite{rem02}.

The pulse profile in the 
0.5--2.0 keV range shows a single, narrow pulse with a high PF~= 
(74$\pm$14)\%.
We then modeled the profile with a 1-D gaussian using $\textit{Sherpa}$ 
(v.3.2.0). The best-fit value for the full-width at
half maximum was 0.26$^{+0.08}_{-0.06}$$P$ (1$\sigma$ errors) with 
$\chi^{2}$(dof) = 2.1(4) and a probability of 0.72. A sinusoidal 
model resulted in $\chi^{2}$(dof) = 8.9(5) with a probability of 
0.11. Due to the limited statistics available, the gaussian 
fit is preferred with a probability of only 2.3$\times$10$^{-2}$ 
(according to the $F$-test). Additional X-ray observations at higher 
resolution are needed to further constrain the pulse shape.

Correlating the times of arrival for the radio pulse at 1.4~GHz with the 
X-ray pulse, we find that the radio peak arrives 26~ms before the centre 
of the X-ray peak (with $\sim$3~ms uncertainty). This corresponds to 
phase 0 in the pulse profiles of Fig. \ref{FigProfPha}. Taking into account 
the low temporal resolution of the \xmm\ observation (phase bin width of 51 
ms), this result suggests that the radio peak is in phase with the X-ray 
peak within our uncertainties, or possibly just slightly ahead.
In addition, although it is unlikely that the EPIC-PN timing anomaly 
present during the observation affected the absolute timing (\xmm\ 
HelpDesk, private communication), we regard this result with caution until 
it can be confirmed. 

\section{Spectral Analysis}\label{SecSpec}
The EPIC data were used to perform a spectral analysis of PSR \psr. 
Circular regions with radii of 20$''$ and 25$''$ were used for MOS and 
PN, encompassing $\sim$75\% and $\sim$78\% of the source photons, 
respectively. The derived fluxes have been corrected accordingly.  
Background regions were chosen from nearby areas away from bright SNR 
knots. The spectra were fit in the 0.5--10.0 keV range using
XSPEC (v.11.3.0) with a minimum of 20 counts per bin from a total of 
240$\pm$19, 210$\pm$18, and 620$\pm$30 background-subtracted counts in 
MOS1, MOS2 and PN, respectively.

%\subsection{Phase-Averaged Spectroscopy}\label{SecAvg}
One-component models were first fit to the phase-averaged MOS and PN data. 
A power-law model provided a better fit than a thermal blackbody model 
(BB), giving $\chi^{2}$(dof) = 113(68) and $\chi^{2}$(dof) = 180(68), 
respectively. However, it was evident from the fit residuals and the 
improvement in $\chi^{2}$ that two-component models were needed in order 
to describe the low and high energy portions of the spectra.
The derived fits from two-component models are summarised in Table 
\ref{tabAvg}. In these fits, a non-thermal power-law component with photon 
index $\Gamma$ $\sim$ 1.5 described the high-energy emission in the spectra 
well. In turn, various models were used to describe the low-energy emission.
A magnetised hydrogen atmosphere model \citep[Atm;][]{zps96} required a 
small distance of $\lesssim$~2~kpc to account for the observed emission (or 
conversely an implausibly large emitting radius of 27$^{+4}_{-2}$~km at 8.4~kpc, 
1$\sigma$ range). Such a small distance can be ruled out from HI absorption
measurements \citep{cmc04}. Currently available (non-magnetic) atmosphere 
models of higher metallicity \citep{gbr02} were also used and found to produce 
poor fits to the observed emission. At a fixed distance 
of 8.4~kpc, the derived temperatures were $\sim$1.9~MK [$\chi^2$(dof)=85(67)] 
and $<$1.2~MK [$\chi^2$(dof)=114(67)] for solar and pure iron compositions, 
respectively. However, when the distance was allowed to be a free parameter, 
the models converged to unrealistic values (e.g., $<$200~pc for temperatures 
$<$1.4~MK) with no improvement on the resulting $\chi^2$. 

We also extracted PN spectra from the ``pulsed'' and ``unpulsed'' regions 
of the pulse profile, at phases 0.7--1.3 (430$\pm$28 counts) and 0.3--0.7 
(200$\pm$20 counts), respectively. These spectra are shown in Figure 
\ref{FigProfPha} (bottom) and were well fit by two-component models that 
agree with those 
derived for the phase-averaged spectra. For example, a blackbody plus power-law
model yielded $T^{\infty}$ = 2.8$\pm$0.4~MK and $\Gamma$ = 1.4$^{+0.5}_{-0.2}$ 
(1$\sigma$ errors). The main difference between the pulsed and unpulsed spectra 
was found to be the relative 
contributions of the model components. The pulsed spectrum is dominated by the 
soft component below 2~keV, where it contributes $\sim$85\% ($\sim$90\%) of the 
absorbed (unabsorbed) flux. In turn, the unpulsed spectrum is dominated by the 
hard, power-law component even below 2~keV, where it contributes $\sim$65\% 
($\sim$55\%) of the absorbed (unabsorbed) flux. For both pulsed and unpulsed 
spectra, the power-law component contributes $>$90\% of the flux above 2~keV 
(either absorbed or unabsorbed).

\section{Discussion}

\subsection{Emission Characteristics}\label{SecEmis}
The \xmm\ observation of PSR \psr\ allowed us to detect, for the first
time, X-ray pulsations from this high-magnetic-field pulsar. The
pulsations are only detected in the 0.5--2.0~keV range with a single, 
narrow peak and a very high pulsed fraction of (74$\pm$14)\%. 

The emission detected within the \xmm\ extraction regions is a 
combination of pulsar, PWN and diffuse SNR emission (see Fig. \ref{FigMOS}). 
In addition, the derived spectra from these regions are well fit by 
two-component models. We then argue that the hard spectral component
detected arises primarily from the pulsar's surroundings and not from the 
NS itself. Using a high-resolution archival \chan\ observation of the field 
\citep[see][]{gs03}, we estimate that the PWN plus SNR emission within a 
$25''$ radius, excluding the pulsar, is well described by a power-law model 
with $\Gamma$=1.8$^{+0.8}_{-0.6}$ and unabsorbed flux in the 0.5--10.0 keV 
range of 0.9$^{+0.6}_{-0.5}$$\times$10$^{-13}$ erg~s$^{-1}$~cm$^{-2}$ 
(1$\sigma$ errors). These values are in good agreement with the hard, 
power-law component shown in Table \ref{tabAvg}. Although a small contribution 
from the pulsar to this hard emission cannot be ruled out, it would not 
affect our results on the pulsar's soft emission.

The soft component detected with \xmm\ must then arise from PSR \psr, as is 
further supported by the pulsed emission having a distinctly soft spectrum (see 
\S\ref{SecSpec}). Such emission is unlikely to have a non-thermal origin.
Non-thermal X-rays from radio pulsars have photon 
indices in the range 0.5$<$$\Gamma$$<$2.7 \citep[e.g.,][]{got03}. 
Synchrotron emission models due to accelerated particles in the 
magnetosphere predict $\Gamma$ $\lesssim$ 2 \citep[e.g.,][]{cz99,rd99}.
The soft component of the pulsed emission from PSR \psr, if interpreted 
as non-thermal, has a steeper spectrum ($\Gamma$=6.5$\pm$0.9) 
than those observed or predicted and would imply an 
acceleration mechanism that cannot be achieved with current 
models. %We then conclude that the soft component has a thermal origin.

The derived blackbody temperature for PSR \psr\ of 
$T_{bb}^{\infty}$~= 2.4$^{+0.3}_{-0.2}$~MK 
is among the highest seen for radio pulsars. 
Although it is similar to those found in a few other pulsars, none of 
them exhibits a higher temperature at statistically 
significant levels \citep[e.g, PSR J0218+4232 has a characteristic age of 
$\tau_{c}$ = $P$/2$\dot{P}$ = 5.1$\times$10$^{8}$~yr and a blackbody 
temperature of $T_{bb}^{\infty}$ = 2.9$\pm$0.4~MK;][]{wob04}. In addition, 
PSR \psr\ is now the youngest radio pulsar from which thermal emission 
has been detected, the next youngest being Vela ($\tau_{c}$ = 11~kyr) 
with a blackbody temperature $T_{bb}^{\infty}$ = 1.47$\pm$0.06~MK 
\citep{pzs+01b}. For comparison, the young Crab pulsar ($\tau_{c}$ = 1.3~kyr) 
and PSR J0205+6449 ($\tau_{c}$ = 5.4~kyr, in the SNR 3C 58) have 
upper limits on the possible blackbody emission from 
their entire surface of $T_{bb}^{\infty}$$<$1.97~MK \citep{wop+04} and 
$T_{bb}^{\infty}$$<$1.02~MK \citep{shvm04}, respectively, at the 3$\sigma$ 
confidence level. 

The observed pulsed fraction for the thermal emission in PSR
\psr\ is also higher than those of other radio pulsars. Sources 
for which the thermal emission arises from the entire surface or from 
localized regions show pulsed fractions at low energies of $<$42\% 
\citep[e.g.,][]{pzs+01b,pzs02,ba02}. The narrow peak in the pulse 
profile also appears to differ from the broad pulsations seen in thermal 
emission from radio pulsars at low energies. Although a multi-component 
pulse profile (e.g., sine curve plus a narrow peak) in the 0.5--2.0 keV 
range cannot be ruled out for PSR \psr\ and requires additional 
observations, we expect any pulse components to show a soft spectrum
consistent with the 3$\sigma$ values in Table \ref{tabAvg} and the 
narrow component to exhibit a high PF $>$ 50\%.

\subsection{Thermal Emission Mechanisms} \label{SecMech}
We first consider the observed characteristics in light of 
conventional models for NS emission. Thermal emission from polar-cap 
reheating, due to return currents from the ``outer gap'' \citep{cz99} 
or from close to the polar cap \citep{hm01b}, is expected to have
an X-ray luminosity of $\lesssim$ 10$^{-5}$$\dot{E}$ in young sources.
This is at least 2 orders of magnitude below our value for the luminosity
of 0.9$^{+1.1}_{-0.2}$$\times$10$^{-3}$$\dot{E}$ (see Table \ref{tabAvg},
BB+PL model). 
In addition, although polar-cap reheating has been proposed to 
explain similar characteristics to that of PSR \psr\ in some 
older pulsars \citep[$\tau_{c}$~$\gtrsim$ 1~Myr, e.g.,][]{ba02}, such 
emission implies much smaller emitting areas \citep[radius $<$ 1~km, 
see e.g.,][]{wob04}. 

Thermal emission may also arise from the entire surface of the NS due to 
initial cooling. While the observed blackbody luminosity is in line with
predictions from cooling models, the observed blackbody temperature is 
slightly higher than predicted \citep[$\lesssim$1.8~MK, e.g.,][]{ygk+04} and
the blackbody radius of $R_{bb}^{\infty}$ = 3.4$^{+4.3}_{-0.9}$~$d_{8.4}$~km 
(3$\sigma$ errors) is smaller than allowed from NS equations of 
state \citep[$R^{\infty}$$\gtrsim$12~km,][]{lp00b}.
The surface temperature can be made more consistent with those 
expected from initial cooling if a hydrogen atmosphere model is used 
\citep[see Table \ref{tabAvg}, and e.g.,][]{pzs02}. However, a recent study 
argues that hydrogen cannot be maintained long-term in atmospheres of NSs 
having magnetic fields as high as that of PSR \psr\ due to diffusive nuclear 
burning, with only heavier elements surviving long-term \citep{cab04}. 
Regardless of this, atmospheric models result in high $\chi^{2}$ values and/or 
problematic estimates for the distance/emitting area (see \S\ref{SecSpec}).

If the thermal emission originates from the entire surface due to initial 
cooling, the pulse profile and pulsed fraction in particular are highly 
problematic, as low amplitude, broad, sinusoidal profiles are expected 
\citep[e.g.,][]{ba02,pzs02}. The surface brightness distribution on a 
highly magnetic star cannot produce pulsed fractions higher than 37\% 
due to General Relativistic light bending. This applies for all 
reasonable NS compactnesses ($\propto$ $M$/$R$, the mass-to-radius ratio 
of the star), extreme viewing geometries and large beaming of the emergent 
radiation \citep{dpn01}. 

Larger pulsed fractions can be obtained if the 
emission arises from localized hot spots on a highly magnetised atmosphere.
However, the possible physical mechanisms leading to this geometry are not 
well understood and has led to suggestions of, e.g., metallicity gradients on 
the surface due to magnetically channelled fallback material \citep{pza+00} 
or nonuniform heating of the atmosphere due to magnetic field decay 
\citep{td96a}. In any case, previous work suggests that if two (isotropic) 
antipodal hot spots are present, the large pulsed fraction in PSR \psr\ 
can only be achieved for extreme viewing geometries, extremely small or 
large NS compactnesses, and small spot sizes 
\citep{pod00,opk01}. These contraints are relaxed if a single hot spot 
is present on the surface, although it is particularly uncertain how 
such a geometry could be physically achieved; perhaps the NS has 
an off-centre magnetic dipole. Regardless of this, accounting for the higher 
effective temperature in PSR \psr\ than in other young pulsars is still 
an issue. 

Currently, modelling of the 
resulting pulsed fraction from hot spots using appropriate atmospheric 
parameters for the case of PSR \psr\ is underway (W. Ho \& 
P. Chang, private communication). Preliminary results suggest that 
improved atmospheric models and including effects from limb darkening
due to magnetic fields can reproduce the observed pulsed fraction over
a wider range of parameters for hot spot emission. However, broad 
pulsations are expected in this case and the physical origin for such 
spots is still uncertain.

Thus, understanding the thermal emission from PSR \psr\ in light of 
conventional NS emission mechanisms requires further theoretical 
studies to be made. Although the particular reason for such emission
is not clear at present, it is likely to be related to the high inferred
magnetic field, an unusual surface temperature distribution and/or an 
unusual beaming from the surface. The pulsar's radio emission 
reveals nothing out of the ordinary that would hint at such peculiar X-ray 
properties. The X-ray characteristics distinguish PSR \psr\ from
other radio pulsars detected at X-ray energies, even from
those with similar spin characteristics (of which there 
are only a few). For example, PSR J1846--0258 in the SNR Kes 
75 has almost identical spin 
properties ($P$ = 0.324~sec, $B$ = 4.8$\times$10$^{13}$~G and age 
$\lesssim$1,000--1,700~yr). However, it is a bright X-ray source 
with a standard power-law spectrum with $\Gamma$~= 1.39$\pm$0.04 and broad 
pulse profile \citep{hcg03,mbb+02}. 

The emission observed from PSR \psr\ also allows us to consider 
alternate models from those of radio pulsars. For 
example, it has been proposed that some (and maybe all) neutron 
stars are quark stars \citep{wit84b}, which are generally thought to 
cool faster than NSs or become indistinguishable from 
them after an initial fast cooling period \citep{web04}. Quark 
stars can be kept hotter for longer by 
including the effects of $e^{+}$$e^{-}$ pair formation from the 
surface, although the observed emission would harden and deviate 
largely from a blackbody \citep{pu02}. This does not appear to be 
the case for PSR \psr\ as its spectrum is well described as a 
blackbody. Newly discovered photon 
emission processes are also expected to keep the star hot and conserve 
the blackbody spectrum \citep{vro04}, but detailed models of the 
cooling history in this case are not currently available. 

Unlike radio pulsars, the observed emission from ``magnetars'' is 
thought to be powered by the decay of an ultrastrong magnetic field 
\citep[10$^{14-15}$~G,][]{td95}. The high thermal temperature and
high pulsed fraction for PSR \psr\ is reminiscent of magnetar emission 
\citep[see, e.g.,][for a review]{wt04}. The high pulsed fraction and 
observed pulse profile for PSR \psr, as well as for the Anomalous X-ray
Pulsars, are consistent with magnetar models where the emission arises 
from a single hot spot on the surface \citep{opk01,oze02}, although, 
again, the reason for this geometry is unclear.
%However, in contrast to what is observed for PSR \psr, 
These models also predict little dependence of the pulsed fraction with 
energy, which can be tested in PSR \psr\ with additional observations. 
However, unlike PSR \psr, magnetars show a significant (pulsed) power-law 
contribution in their 
spectra, have X-ray luminosities much higher than the available rotational 
energy and, thus far, have not shown radio emission. To date, no other 
apparently high magnetic field radio pulsar has shown emission characteristics
similar to that of magnetars (e.g., PSR J1814--1744, Pivovaroff et al. 2000;
PSR J1846--0258, Helfand et al. 2003; PSR J1718--3718, Kaspi \& McLaughlin 
2004), although it has been suggested that they represent magnetars in 
quiescence \citep{km04}. 

%Finally, the observed thermal emission from PSR \psr\ with high temperature, 
%narrow pulse profile, high pulsed fraction and inferred emission radius is 
%unlike those of other young neutron stars. 
%Given the unusual properties of the emission, deeper X-ray 
%observations %carried out at higher temporal resolution
%are needed to further constrain the emission characteristics. As well,
%additional detailed studies of NS atmosphere emission models
%are needed to confront with these observations.

\acknowledgements
We thank A. Cumming, A. Harding, R. Ouyed, M. Roberts, L. Bildstein, D. 
Lai, W. Ho, P. Chang and R. Rutledge for useful discussions. This work 
was supported by 
an NSERC Research Grant and NASA grant NAG5-13572. VMK is a Canada 
Research Chair and Steacie Fellow and received support from an NSERC 
Discovery Grant and Steacie Supplement, from FQRNT and CIAR.
Radio timing data for PSR \psr\ were collected at 
the Parkes telescope, a part of the Australia Telescope which
is funded by the Commonwealth of Australia for operation
as a National Facility managed by CSIRO.  We thank the
members of the multibeam survey team, in particular
R. Manchester, for help with collecting these data.

\bibliography{journals1,psrrefs,modrefs}

%\clearpage

\begin{deluxetable}{lccc}
%\rotate 
\tabletypesize{\footnotesize} 
\tablewidth{0pt}
\tablecaption{\label{tabAvg} Fits to the \xmm\ phase-averaged spectrum of PSR \psr} 
\tablehead{ \colhead{} & \colhead{PL + PL} & \colhead{BB + PL} & 
\colhead{Atm\tablenotemark{a} + PL} \\ 
\colhead{Parameter} & \colhead{($\pm$1$\sigma$)} & 
\colhead{($\pm$1$\sigma$)} & \colhead{($\pm$1$\sigma$)}}
\startdata
$N_{H}$ (10$^{22}$~cm$^{-2}$) & 2.3$^{+0.4}_{-0.3}$ & 1.6$^{+0.4}_{-0.3}$ & 1.9$^{+0.5}_{-0.3}$ \\
$\chi^2$(dof) & 79(66) & 78(66) & 78(66) \\ %\hline
& \multicolumn{3}{c}{\it Soft component characteristics} \\ %\hline
$\Gamma$ or $T^{\infty}$ & 6.5$\pm$0.9 & 2.4$^{+0.3}_{-0.2}$~MK &  0.9$\pm$0.2~MK \\
$R^{\infty}$ (km) & ... & 3.4$^{+1.8}_{-0.3}$ & 12 (fixed) \\
$d$ (kpc) & 8.4 (fixed) & 8.4 (fixed) & 1.6$^{+0.2}_{-0.9}$ \\
$f_{abs}$ \tablenotemark{b} (10$^{-14}$) & 2.1$^{+2.3}_{-0.9}$ & 1.5$^{+1.8}_{-0.2}$ & 1.7$^{+7.0}_{-0.4}$ \\
$f_{unabs}$ \tablenotemark{b} (10$^{-13}$) & 63$^{+57}_{-32}$ & 2.4$^{+3.0}_{-0.5}$ & 7.2$^{+31}_{-1.6}$ \\
$L_{X}$ \tablenotemark{b} (10$^{33}$) & 53$^{+50}_{-27}$ & 2.0$^{+2.5}_{-0.4}$ & 0.22$^{+0.88}_{-0.05}$\\  %\hline
& \multicolumn{3}{c}{\it Hard component characteristics} \\ %\hline
$\Gamma$ & 1.3$^{+0.5}_{-0.2}$ & 1.5$^{+0.3}_{-0.2}$   & 1.5$^{+0.2}_{-0.3}$ \\
$f_{abs}$ \tablenotemark{b} (10$^{-14}$) & 7.1$^{+10}_{-1.5}$   & 7.4$^{+3.6}_{-1.0}$ & 7.3$^{+4.7}_{-2.7}$ \\
$f_{unabs}$ \tablenotemark{b} (10$^{-13}$) & 1.0$^{+1.6}_{-0.2}$  & 1.1$^{+0.6}_{-0.2}$ & 1.1$^{+0.8}_{-0.3}$ \\
$L_{X}$ \tablenotemark{b} (10$^{33}$) & 0.8$^{+1.3}_{-0.2}$ & 0.9$^{+0.5}_{-0.1}$ & 0.04$\pm$0.02 \\ 
\enddata
%\caption{\label{table}}
\tablenotetext{a}{ The atmospheric model was computed with $B$=10$^{13}$~G and pure 
hydrogen composition. The local values for the temperature, $T$, and radius, 
$R$ = 10~km, of the star have been redshifted to infinity according to $T^{\infty}$ 
= $T$(1$-$2$G$$M$/$R$$c^{2}$)$^{1/2}$ and $R^{\infty}$ = 
$R$(1$-$2$G$$M$/$R$$c^{2}$)$^{-1/2}$, with $M$ = 1.4~$M_{\sun}$.}
%\tablenotetext{b}{ The blackbody radius is proportional to the distance.}
\tablenotetext{b}{ The 0.5--10.0 keV absorbed and unabsorbed fluxes, $f_{abs}$ 
and $f_{unabs}$, have units of ergs~s$^{-1}$~cm$^{-2}$. The 0.5--10.0 keV X-ray 
luminosity, $L_X$, at the distance $d$, is in units of ergs~s$^{-1}$.}
\end{deluxetable}

\end{document}